\documentclass[prb,twocolumn,superscriptaddress,longbibliography]{revtex4-1}
\RequirePackage[mathlines]{lineno}
\usepackage{amssymb}
\usepackage{amsmath}
\usepackage{stmaryrd}
\usepackage{mathrsfs}
\usepackage{graphicx}
\usepackage[usenames,dvipsnames]{xcolor}
\definecolor{goodgreen}{rgb}{0.1,0.5,0}
\definecolor{goodred}{rgb}{0.7,0,0}
\usepackage{natbib}
\usepackage[colorlinks,urlcolor=goodgreen,citecolor=blue,linkcolor=goodred]{hyperref}

\usepackage{tikz}
\usetikzlibrary{calc}
\usetikzlibrary{shapes.geometric, arrows.meta, positioning}
\usetikzlibrary{positioning}

\makeatletter
\newsavebox{\@brx}
\newcommand{\llangle}[1][]{\savebox{\@brx}{\(\m@th{#1\langle}\)}%
  \mathopen{\copy\@brx\kern-0.5\wd\@brx\usebox{\@brx}}}
\newcommand{\rrangle}[1][]{\savebox{\@brx}{\(\m@th{#1\rangle}\)}%
  \mathclose{\copy\@brx\kern-0.5\wd\@brx\usebox{\@brx}}}
\makeatother

\usepackage[normalem]{ulem}

\setpagewiselinenumbers
\modulolinenumbers[5]
\begin{document}

%\preprint{\dcom{Preprint not for distribution CONFIDENTIAL, Version of \today}}
\title{Stacking-Induced Large-Chern-Number Quantum Anomalous Hall Phases}

\author{H. Minh Lam}
\author{V. Nam Do}
\email{nam.dovan@phenikaa-uni.edu.vn}    
\affiliation{Phenikaa Institute for Advanced Study (PIAS), A9 Building, Phenikaa University, Hanoi 12116, Vietnam}

\begin{abstract}
We investigate the interaction between quantum anomalous Hall (QAH) phases hosted by two atomically thin hexagonal lattices and demonstrate the emergence of topological phases with large Chern numbers. Interlayer coupling between two graphene-like lattices produces band crossings, while relative sliding preserves gapless Dirac points located at generic, low-symmetry $\mathbf{k}$ points. The introduction of Haldane-type complex next-nearest-neighbor hoppings gaps these Dirac points, breaks time-reversal symmetry, and generates a sequence of quantum anomalous Hall (QAH) phases. Depending on the phase angles $\phi_1$ and $\phi_2$ associated with the two layers, the system exhibits QAH states with Chern numbers $|C|>2$. The nontrivial bulk topology is verified by the presence of the corresponding number of chiral edge modes in ribbon geometries. These high–Chern-number phases originate from the enhanced twisting of the valence-band manifold induced by interlayer stacking.
\end{abstract}
\maketitle

\section{Introduction}\label{Sec_I}
Chern insulators are realized not only as genuine electronic phases in two-dimensional crystalline systems but also as prototypical models for a broad class of topological phases. Characterized by chiral edge states protected by topological invariants, finite-size Chern insulators hold promise for spintronic and low-dissipation electronic applications.\cite{Moore_2010,Qi_2011,Ren_2016,Tokura_2019,Breunig_2022,Liu_2023} Their potential use in nanoscale interconnects has attracted particular interest, as chiral edge modes enable quantized Hall conductance with negligible backscattering. A major obstacle to practical implementation, however, is the large contact resistance at metal–topological-insulator interfaces. Because this resistance decreases inversely with the number of conducting edge channels, realizing phases with Chern numbers higher than 2 has become a central goal.\cite{Qi_2011,Ren_2016,Zhang_2012,Zhao_2020} Achieving such phases, nevertheless, remains challenging.

Recent theoretical and experimental studies have explored various routes toward realizing QAH phases with large Chern numbers. A central idea is to enhance the twisting of the valence manifold by introducing multiple or higher-order band-touching points between the valence and conduction bands. This can be achieved by increasing the number of internal degrees of freedom (spin and orbitals) and introducing couplings among them, for example, by (i) incorporating longer-range hopping processes,\cite{Sticlet_2013,Yang_2012,Woo_2024} (ii) utilizing multiple orbitals per unit cell,\cite{Trescher_2012,Cook_2016,Alase_2021} (iii) employing twisted bilayer structures,\cite{Wang_2024,Chen_2020,Liu_2019,Khosravian_2024} (iv) constructing moir{\'e} superlattices,\cite{Wang_2021,Fang_2014,Zhao_2020} or (v) stacking multiple magnetically doped or intrinsically magnetic topological insulator layers.\cite{Fang_2014,Wang_2021,Zhao_2020,Li_2022,Zhu_2022,Bosnar_2023,Li_2025} 
Despite these advances, most engineering strategies encounter intrinsic limitations. Stacking topological insulator layers separated by buffer layers generally yields only the additive sum of the Chern numbers of the individual layers,\cite{Wang_2021,Jiang_2012,Zhu_2022,Ge_2020,Li_2022,Wang_2023,Xiong_2016} whereas approaches relying on strong spin--orbit coupling typically require additional mechanisms to break time-reversal symmetry. Models based on long-range hopping are challenging to realize experimentally, and flat-band designs involving higher-order band-touching points generally result in fragile topological phases that are highly sensitive to disorder, symmetry breaking, band hybridization or mixing, and other perturbations.

This work investigates the direct coupling between QAH phases, with the aim of elucidating how such phases are preserved, suppressed, or transformed into new topological states with large Chern numbers in experimentally accessible systems. Specifically, we focus on bilayer configurations formed by stacking two atomically thin hexagonal lattices, exemplified by graphene bilayers. Due to the high symmetry of each lattice, Dirac cones appear at the high-symmetry $\mathbf{K}$ points, which are typically hybridized by interlayer coupling. Previous studies have primarily considered simplified models that include only vertical interlayer hoppings, as in Bernal-stacked configurations.~\cite{Bhattacharjee_2021,Mondal_2023} Here, we extend this picture by explicitly incorporating skew interlayer hoppings, providing a more realistic description of the interlayer coupling landscape. We find that sliding the layers alone does not open a global energy gap but instead induces both vertical (energy) and lateral (momentum-space) displacements of the bands, resulting in crossing and touching points at generic, low-symmetry $\mathbf{k}$ positions. Introducing Haldane-type complex next-nearest-neighbor hoppings gaps these Dirac points, leading to band inversion between the valence and conduction manifolds. The ensuing twisting of the valence manifold generates nonzero Chern numbers. In the parameter space spanned by the two magnetic phases $(\phi_1,\phi_2)$, we identify regions with Chern numbers of magnitude greater than two ($|C|>2$), reflecting the nontrivial interplay between the topological characters of the two layers rather than a simple sum of their individual Chern numbers.~\cite{Bhattacharjee_2021,Mondal_2023} The presence of these high–Chern-number phases is further corroborated by edge-state calculations in ribbon geometries, which reveal three pairs of chiral edge modes, consistent with the bulk Chern numbers.

It is worth emphasizing that, in these bilayer systems, the interlayer coupling is governed not only by the layer separation but also by their relative lateral displacement (sliding). This provides a practical route for realizing high–Chern-number QAH phases, as the twisting of the valence-band manifold can be effectively controlled through the interlayer coupling in atomically stacked lattices.

The remainder of this paper is organized as follows. In Sec.~\ref{Sec_II}, we introduce the bilayer model and computational methods. In Sec.~\ref{Sec_III}, we present the topological phase diagrams and band-structure analysis. Finally, Sec.~\ref{Sec_IV} summarizes our findings and outlines possible directions for future work.

\section{Model and Calculation Method}\label{Sec_II}
Graphene is a prototypical two-level system with a honeycomb lattice and $P6mm$ point-group symmetry.\cite{Do_2021} Its electronic band structure hosts two inequivalent Dirac points at the $\mathbf{K}$ and $\mathbf{K}^\prime$ points of the Brillouin zone, where the quantum-state phase becomes ill-defined due to gauge ambiguity, producing singularities in the mapping from momentum space to the Bloch sphere. These singularities enable topological phenomena; however, the two Dirac points are related by time-reversal symmetry and stabilized by threefold rotational symmetry, leading to equal and opposite Berry curvature contributions that cancel. Thus, pristine graphene is topologically trivial in the Chern classification.

The Haldane model circumvents this limitation by introducing complex next-nearest-neighbor hopping that explicitly breaks time-reversal symmetry, generating mass gaps of opposite sign at the two Dirac points and producing a fully gapped band structure with a nonzero Chern number.\cite{Haldane_1988} This mechanism underlies the quantum anomalous Hall effect. Below, we present a model for sliding bilayer configurations by specifying their lattice geometries and interlayer couplings.

\subsection{Geometry of Atomic Lattices}
We consider bilayer systems composed of two hexagonal lattices without interlayer twist, where the relative in-plane displacement between layers is characterized by a sliding vector $\boldsymbol{\tau}$. The configuration with $\boldsymbol{\tau} = \mathbf{0}$ corresponds to the AA-stacked bilayer. For $\boldsymbol{\tau} \ne \mathbf{0}$, we refer to the system as a sliding bilayer (SBL) configuration.

To define the atomic geometry, we adopt a Cartesian coordinate system $Oxyz$ with $\hat{\mathbf{x}}$ and $\hat{\mathbf{y}}$ spanning the plane of the layers and $\hat{\mathbf{z}}$ oriented normal to the plane. The sliding operation preserves the planarity and translational symmetry of the bilayer, consistent with the AA-stacked lattice. The primitive lattice vectors are given by
\begin{equation}
\mathbf{a}_1 = \frac{a}{2}(\sqrt{3}\,\hat{\mathbf{x}} + \hat{\mathbf{y}}), \quad
\mathbf{a}_2 = \frac{a}{2}(\sqrt{3}\,\hat{\mathbf{x}} - \hat{\mathbf{y}}), \label{Eq1}
\end{equation}
where $a = \sqrt{3}a_{CC}$ is the lattice constant, and $a_{CC}$ is the nearest-neighbor carbon-carbon bond length.

The reciprocal lattice vectors are
\begin{equation}
\mathbf{b}_1 = \frac{2\pi}{\sqrt{3}a}(\hat{\mathbf{x}} + \sqrt{3}\,\hat{\mathbf{y}}), \quad
\mathbf{b}_2 = \frac{2\pi}{\sqrt{3}a}(\hat{\mathbf{x}} - \sqrt{3}\,\hat{\mathbf{y}}), \label{Eq2}
\end{equation}
satisfying $\mathbf{a}_i \cdot \mathbf{b}_j = 2\pi\delta_{ij}$. The high-symmetry $\mathbf{K}$ and $\mathbf{K}^\prime$ points in the Brillouin zone are located at
\begin{equation}
\mathbf{K} = \frac{2}{3}\mathbf{b}_1 + \frac{1}{3}\mathbf{b}_2, \quad
\mathbf{K}' = \frac{1}{3}\mathbf{b}_1 + \frac{2}{3}\mathbf{b}_2. \label{Eq3}
\end{equation}

Each unit cell contains four carbon atoms: $A_1$ and $B_1$ in the bottom layer, and $A_2$ and $B_2$ in the top layer. The choice of origin $O$ is important for consistent atomic positioning across different $\boldsymbol{\tau}$ values. For all SL configurations except AA stacking, the origin is chosen at the inversion center located at the midpoint of the quadrilateral formed by atoms $A_1$, $A_2$, $B_2$, and $B_1$. In the AA-stacked case ($\boldsymbol{\tau} = \mathbf{0}$), the origin is placed at the center of a hexagonal plaquette.

Letting $\mathbf{r}_1 = (\mathbf{a}_1 + \mathbf{a}_2)/3$ and $d_{GG}$ denote the interlayer spacing, the atomic positions in the unit cell are
\begin{subequations}\label{Eq4}
\begin{align}
\mathbf{r}_{A_1} &= +\tfrac{1}{2}(\boldsymbol{\tau} + d_{GG}\,\hat{\mathbf{z}} - \mathbf{r}_1), \label{Eq4a} \\
\mathbf{r}_{B_1} &= +\tfrac{1}{2}(\boldsymbol{\tau} + d_{GG}\,\hat{\mathbf{z}} + \mathbf{r}_1). \label{Eq4b} \\
\mathbf{r}_{A_2} &= -\tfrac{1}{2}(\boldsymbol{\tau} + d_{GG}\,\hat{\mathbf{z}} + \mathbf{r}_1), \label{Eq4c} \\
\mathbf{r}_{B_2} &= -\tfrac{1}{2}(\boldsymbol{\tau} + d_{GG}\,\hat{\mathbf{z}} - \mathbf{r}_1), \label{Eq4d}
\end{align}
\end{subequations}

To classify SBL configurations, we introduce two auxiliary vectors
\[
\boldsymbol{\tau}_1 = \tfrac{1}{3}(\mathbf{a}_1 + \mathbf{a}_2), \quad 
\boldsymbol{\tau}_2 = \tfrac{1}{2}(\mathbf{a}_1 - \mathbf{a}_2),
\]
which are oriented along the $x$ and $y$ directions, respectively. The sliding vector can then be expressed as
\[
\boldsymbol{\tau} = \xi \boldsymbol{\tau}_1 + \zeta \boldsymbol{\tau}_2 \equiv (\xi, \zeta).
\]
Based on the symmetry of the system, the SL configurations can be grouped into five symmetry classes as summarized in Table~\ref{tab:configurations}.

\begin{table}[h]
\centering
\caption{Classification of sliding bilayer configurations by symmetry and geometry.}
\small
\renewcommand{\arraystretch}{1.2}
\begin{tabular}{c|c|c|p{2.24cm}}
\hline\hline
Configs. & $\boldsymbol{\tau}$ & Sym. group & Remarks \\
\hline
C1 & $(0,0)$ & $D_{6h}$ & AA-stacked \\
C2 & $(\xi, 0)$ with $0<\xi <1$ & $C_{2v}$ & Sliding along $Ox$ \\
C3 & $(1,0)$ & $D_{3d}$ & AB-stacked \\
C4 & $(0, \zeta)$ with $0<\zeta<1$ & $C_{2h}$ & Sliding along $Oy$ \\
C5 & $(0,1)$ & $D_{3d}$ & Bernal-stacked \\
\hline
\end{tabular}
\label{tab:configurations}
\end{table}

\subsection{Tight-binding model for bulk configurations}

To describe the electronic properties of the SL lattice, we assume that each atom contributes a single valence electron occupying a $p_z$-like orbital. Let $|p_z, \alpha, \mathbf{r}_\alpha, \mathbf{R}\rangle$ denote the orbital centered at atom $\alpha$ ($\alpha = A_1, B_1, A_2, B_2$), located at position $\mathbf{r}_\alpha$ within the unit cell labeled by the Bravais lattice vector $\mathbf{R}$. We define the basis set as
\begin{equation}
    \mathcal{B} = \left\{|p_z, \alpha, \mathbf{r}_\alpha, \mathbf{R}\rangle \,\middle|\, \alpha = A_1, B_1, A_2, B_2;\, \mathbf{R} \in \Gamma \right\},
\end{equation}
where $\Gamma$ denotes the two-dimensional Bravais lattice. Within this basis, the tight-binding Hamiltonian is written as
\begin{equation}
    \hat{H} = \hat{H}_{\mathrm{BL}} + \hat{H}_H,
\end{equation}
where $\hat{H}_{\mathrm{BL}}$ describes intra- and interlayer hopping processes, and $\hat{H}_H$ introduces the complex next-nearest-neighbor Haldane term that breaks time-reversal symmetry.

The hopping Hamiltonian $\hat{H}_{\mathrm{BL}}$ takes the form
\begin{equation}\label{Eq5}
\hat{H}_{\mathrm{BL}} = \sum_{\mathbf{R}} \sum_{j,\alpha,\beta} t_{\alpha\beta}^j\, |p_z, \alpha, \mathbf{r}_\alpha, \mathbf{R}\rangle \langle p_z, \beta, \mathbf{r}_\beta, \mathbf{R}_j|,
\end{equation}
where $t_{\alpha\beta}^j$ is the hopping integral between orbitals centered at $\mathbf{R} + \mathbf{r}_\alpha$ and $\mathbf{R}_j + \mathbf{r}_\beta$. The Haldane term is written as $\hat{H}_H = \sum_{\ell = 1}^{2} \hat{H}_H^\ell$, where each $\hat{H}_H^\ell$ is layer-resolved and given by
\begin{align}\label{Eq14}
\hat{H}_H^\ell &= t_H \sum_{\mathbf{R}} \sum_{\alpha, j=1}^6 e^{i\phi_{\ell n_\alpha j}}\, |p_z, \alpha, \mathbf{r}_\alpha, \mathbf{R}\rangle \langle p_z, \alpha, \mathbf{r}_\alpha, \mathbf{R} + \mathbf{R}_j| \nonumber \\
&\quad + \sum_{\mathbf{R}} \sum_{\alpha} M_{\ell\alpha} \,|p_z, \alpha, \mathbf{r}_\alpha, \mathbf{R}\rangle \langle p_z, \alpha, \mathbf{r}_\alpha, \mathbf{R}|.
\end{align}
The vectors are defined as $\mathbf{R}_1 = -\mathbf{R}_4 = \mathbf{a}_1$, $\mathbf{R}_2 = -\mathbf{R}_5 = \mathbf{a}_1 - \mathbf{a}_2$, and $\mathbf{R}_3 = -\mathbf{R}_6 = -\mathbf{a}_2$. The on-site potentials satisfy $M_{\ell A} = -M_{\ell B} = M_\ell$, and the complex hopping phases are $\phi_{\ell n_\alpha j} = (-1)^{\text{mod}(j,2)}(-1)^{n_\alpha} \phi_\ell$, with $n_\alpha = 1,2$ for $\alpha = A,B$, respectively. This construction induces a net flux with left-handed chirality across each hexagonal plaquette.

To model realistic electronic structures, the hopping amplitudes $t_{\alpha\beta}^j$ are not treated as empirical parameters but are instead evaluated using the Slater–Koster formalism. In this approach, the hopping is determined by two-center integrals depending on the bond vector:
\begin{equation}\label{Eq6}
t(\mathbf{r}_{\alpha\beta}^j) = V_{pp\pi}(r^j_{\alpha\beta}) \sin^2\theta^z_{\alpha\beta} + V_{pp\sigma}(r^j_{\alpha\beta}) \cos^2\theta^z_{\alpha\beta},
\end{equation}
where $\cos\theta^z_{\alpha\beta} = (\mathbf{r}^j_{\alpha\beta} \cdot \mathbf{e}_z)/r^j_{\alpha\beta}$ and $r^j_{\alpha\beta} = \|\mathbf{R}_j + \mathbf{r}_\beta - (\mathbf{R} + \mathbf{r}_\alpha)\|$. The distance-dependent transfer integrals are given by
\begin{subequations}\label{Eq7}
\begin{align}
V_{pp\pi}(r^j_{\alpha\beta}) &= V^0_{pp\pi} \exp\left(-\frac{r^j_{\alpha\beta} - a_{CC}}{r_0}\right), \\
V_{pp\sigma}(r^j_{\alpha\beta}) &= V^0_{pp\sigma} \exp\left(-\frac{r^j_{\alpha\beta} - d_{GG}}{r_0}\right),
\end{align}
\end{subequations}
where $r_0 = 0.184a$, a standard value for graphene-like systems.~\cite{Moon_2013,Le_2018,Lam_2023} The energy scale is set by $V^0_{pp\pi}$, while $V^0_{pp\sigma}$ controls the strength of interlayer coupling.

To obtain the energy band structure, we begin by introducing the Bloch basis states $\{|p_z,\alpha,\mathbf{k}\rangle\}$ as the Fourier transforms of the localized orbitals $\{|p_z,\alpha,\mathbf{r}_\alpha,\mathbf{R}_j\rangle\}$, defined as
\begin{equation}
|p_z, \alpha, \mathbf{r}_\alpha, \mathbf{R}_j\rangle = \frac{1}{\sqrt{N}} \sum_{\mathbf{k}} e^{-i\mathbf{k} \cdot (\mathbf{R}_j + \mathbf{r}_\alpha)} |p_z, \alpha, \mathbf{k}\rangle,
\label{Eq8}
\end{equation}
where $\mathbf{k}$ lies in the rhombohedral region $T^2$ generated by the reciprocal lattice vectors $(\mathbf{b}_1,\mathbf{b}_2)$. Upon substituting Eq.~\eqref{Eq8} into Eq.~\eqref{Eq5}, the bilayer Hamiltonian takes the form
\begin{equation}\label{Eq9}
\hat{H}_{\mathrm{BL}} = \sum_{\mathbf{k}} \sum_{\alpha, \beta} |p_z, \alpha, \mathbf{k}\rangle h_{\alpha\beta}(\mathbf{k}) \langle p_z, \beta, \mathbf{k}|,
\end{equation}
with 
\begin{equation}\label{Eq10}
h_{\alpha\beta}(\mathbf{k}) = \sum_{j} t(\mathbf{r}^j_{\alpha\beta}) e^{-i\mathbf{k} \cdot \mathbf{r}^j_{\alpha\beta}}.
\end{equation}
As shown in Eq.~\eqref{Eq9}, when expressed in the Bloch-state basis $\{|p_z, \alpha, \mathbf{k}\rangle\}$, the Hamiltonian $\hat{H}_{\mathrm{BL}}$ decouples into independent $\mathbf{k}$-sector blocks, hereafter referred to as block Hamiltonian matrices, which take the form
\begin{equation}\label{Eq11}
H_{\mathrm{BL}}(\mathbf{k}) = \begin{pmatrix}
0 & f_\mathbf{k} & u_\mathbf{k} & v_\mathbf{k} \\
f^*_\mathbf{k} & 0 & w_\mathbf{k} & u_\mathbf{k} \\
u^*_\mathbf{k} & w^*_\mathbf{k} & 0 & f_\mathbf{k} \\
v^*_\mathbf{k} & u^*_\mathbf{k} & f^*_\mathbf{k} & 0
\end{pmatrix},
\end{equation}
where the matrix elements are defined as
\begin{subequations}
\begin{align}
f_\mathbf{k} &= \sum_{j} t_f e^{-i\mathbf{k} \cdot \mathbf{r}^j_{A_1B_1}}, \label{Eq12a}\\
u_\mathbf{k} &= \sum_{j} t_u e^{-i\mathbf{k} \cdot \mathbf{r}^j_{A_1A_2}}, \label{Eq12b}\\
v_\mathbf{k} &= \sum_{j} t_v e^{-i\mathbf{k} \cdot \mathbf{r}^j_{A_1B_2}}, \label{Eq12c}\\
w_\mathbf{k} &= \sum_{j} t_w e^{-i\mathbf{k} \cdot \mathbf{r}^j_{B_1A_2}}. \label{Eq12d}
\end{align}
\end{subequations}
with $t_f=t(\mathbf{r}^j_{A_1B_1})$, $t_u=t(\mathbf{r}^j_{A_1A_2})$, $t_v=t(\mathbf{r}^j_{A_1B_2})$ and $t_w=t(\mathbf{r}^j_{B_1A_2})$.

For nearest-neighbor intralayer hopping, Eq.~\eqref{Eq12a} reduces to
\begin{equation}\label{Eq13}
f_\mathbf{k} = t_0 \left[e^{-ik_x a_{CC}} + 2e^{i \frac{k_x a_{CC}}{2}} \cos\left(\frac{\sqrt{3}k_y a_{CC}}{2}\right)\right],
\end{equation}
with $t_0 = V^0_{pp\pi}$. Interlayer hopping is restricted to atomic sites within a cutoff radius $R^c = \sqrt{a_{CC}^2+d_{GG}^2}$.

Transforming the Haldane term into the Bloch basis, the full Bloch Hamiltonian becomes
\begin{align}\label{Eq15}
H(\mathbf{k}) &= H_{\mathrm{BL}}(\mathbf{k}) + 2t_2 \left[\sum_{j=1}^3 \cos(\mathbf{k} \cdot \mathbf{R}_j)\right] \gamma_{\cos} \nonumber \\
&\quad - 2t_2 \left[\sum_{j=1}^3 \sin(\mathbf{k} \cdot \mathbf{R}_j)\right] \gamma_{\sin} + M \sigma_0 \otimes \tau_z,
\end{align}
where $\mathbf{R}_1 = \mathbf{a}_1$, $\mathbf{R}_2 = -\mathbf{a}_2$, and $\mathbf{R}_3 = -\mathbf{a}_1 + \mathbf{a}_2$, with
\begin{subequations}\label{Eq16}
\begin{align}
\gamma_{\cos} &= \text{diag}(\cos\phi_1, \cos\phi_1, \cos\phi_2, \cos\phi_2), \\
\gamma_{\sin} &= \text{diag}(\sin\phi_1, -\sin\phi_1, \sin\phi_2, -\sin\phi_2).
\end{align}
\end{subequations}

It is worth noting that the Bloch basis defined in Eq.~\eqref{Eq8} is convenient for constructing the block Hamiltonian matrices. However, the resulting Hamiltonian $H(\mathbf{k})$ is not periodic under reciprocal-lattice translations $\mathbf{G}=m\mathbf{b}_1+n\mathbf{b}_2$. To analyze topological properties of the Bloch bands, one must work with a Hamiltonian that satisfies lattice periodicity. This can be achieved by performing a unitary gauge transformation,
\[
H(\mathbf{k}) \;\rightarrow\; U(\mathbf{k})\,H(\mathbf{k})\,U^\dagger(\mathbf{k}),
\]
where
\[
U(\mathbf{k}) = \mathrm{diag}\!\left(
e^{-i\mathbf{k}\cdot\mathbf{r}_{A_1}},
e^{-i\mathbf{k}\cdot\mathbf{r}_{B_1}},
e^{-i\mathbf{k}\cdot\mathbf{r}_{A_2}},
e^{-i\mathbf{k}\cdot\mathbf{r}_{B_2}}
\right).
\]
The transformed Hamiltonian $H(\mathbf{k})$ is periodic in the torus $T^2$ and therefore suitable for computing Berry connections, Berry curvature, and Chern numbers.

As shown in Ref.~\onlinecite{Do_2021}, in the absence of time-reversal-symmetry breaking, the SBL lattices exhibits symmetry-protected band crossings enforced by inversion and mirror symmetries. Introducing the Haldane terms on each layer explicitly breaks time-reversal and mirror symmetries, lifting these symmetry-protected degeneracies and giving rise to nontrivial topological phases, as discussed in the following section.

\subsection{Model construction for ribbon configurations}

To verify the bulk-boundary correspondence, we calculate the electronic structure of ribbon configurations with two typical edge types: zigzag and armchair. To maintain consistency with the bulk systems, we construct the armchair-edge (AE) and zigzag-edge (ZE) ribbons by constraining their widths along the $Oy$ and $Ox$ directions, respectively. 

To define the atomic lattice of a ribbon with two atomic layers, we begin by constructing a single-layer ribbon. The second layer is then generated by duplicating the first and stacking it with a relative sliding vector $\boldsymbol{\tau}$. Denoting $N_\text{zig}$ and $N_\text{arm}$ as the number of zigzag and armchair lines in the ribbons, respectively, the ribbon widths are given by
\begin{subequations}\label{Eq17}
    \begin{align}
        W_A &= (N_\text{arm}-1)\frac{\sqrt{3}a_{CC}}{2}, \label{Eq17a}\\
        W_Z &= (3N_\text{zig}-2)\frac{a_{CC}}{2}. \label{Eq17b}
    \end{align}
\end{subequations}
For bilayer ribbons, the sliding vector $\boldsymbol{\tau}$ must be considered to determine the total width.

In AE ribbons, the atomic lattice is periodic along the $Ox$ direction. We define a unit cell as a segment of a zigzag line consisting of $N_A=2N_\text{arm}$ carbon atoms per layer. Each atom is indexed by a pair $(L,j)$, where $j = 1, 2, \dots, 2N_A$, and its position within unit cell $L$ is specified by the vector $\mathbf{r}^j = (d^j_x, d^j_y)$. In ZE ribbons, the atomic lattice is periodic along the $Oy$ direction. A unit cell is taken as a segment of an armchair line with $N_Z=2N_\text{zig}$ carbon atoms per layer. Each atom is indexed by $(L,j)$ with $j = 1, 2, \dots, 2N_Z$.

The $p_z$ orbital of each carbon atom in the AE and ZE ribbons is denoted by $|p_z,\mathbf{r}^j,La_\nu\rangle$, where $a_\nu = \sqrt{3}a$ for $\nu = x$ and $a_\nu = a$ for $\nu = y$. The corresponding Bloch state vectors $|p_z,j,k_\nu\rangle$ are defined as
\begin{equation}\label{Eq20}
|p_z,j,k_\nu\rangle = \frac{1}{\sqrt{N}} \sum_{L} e^{ik_\nu(La_\nu + d_\nu^j)} |p_z,\mathbf{r}^j,La_\nu\rangle,
\end{equation}
where $k_\nu \equiv k_x$ in the range $[-\pi/\sqrt{3}a,\pi/\sqrt{3}a]$ ($\text{BZ}_A$) and $k_\nu \equiv k_y$ in the range $[-\pi/a, \pi/a]$ ($\text{BZ}_Z$). Using the basis $\{|p_z,j,k_\nu\rangle\mid j = 1,2,\dots,2N_{A/Z},\, k_\nu \in \text{BZ}_{A/Z}\}$, we construct the Bloch Hamiltonian matrix $h^{A/Z}(k_\nu)$ for the AE/ZE ribbons. The $2N_{A/Z} \times 2N_{A/Z}$ Hamiltonians are then numerically diagonalized to obtain the ribbon band structures.

\begin{figure*}
    \centering
    \includegraphics[clip,trim=2.8cm 3cm 4.8cm 2.3cm,width=\textwidth]{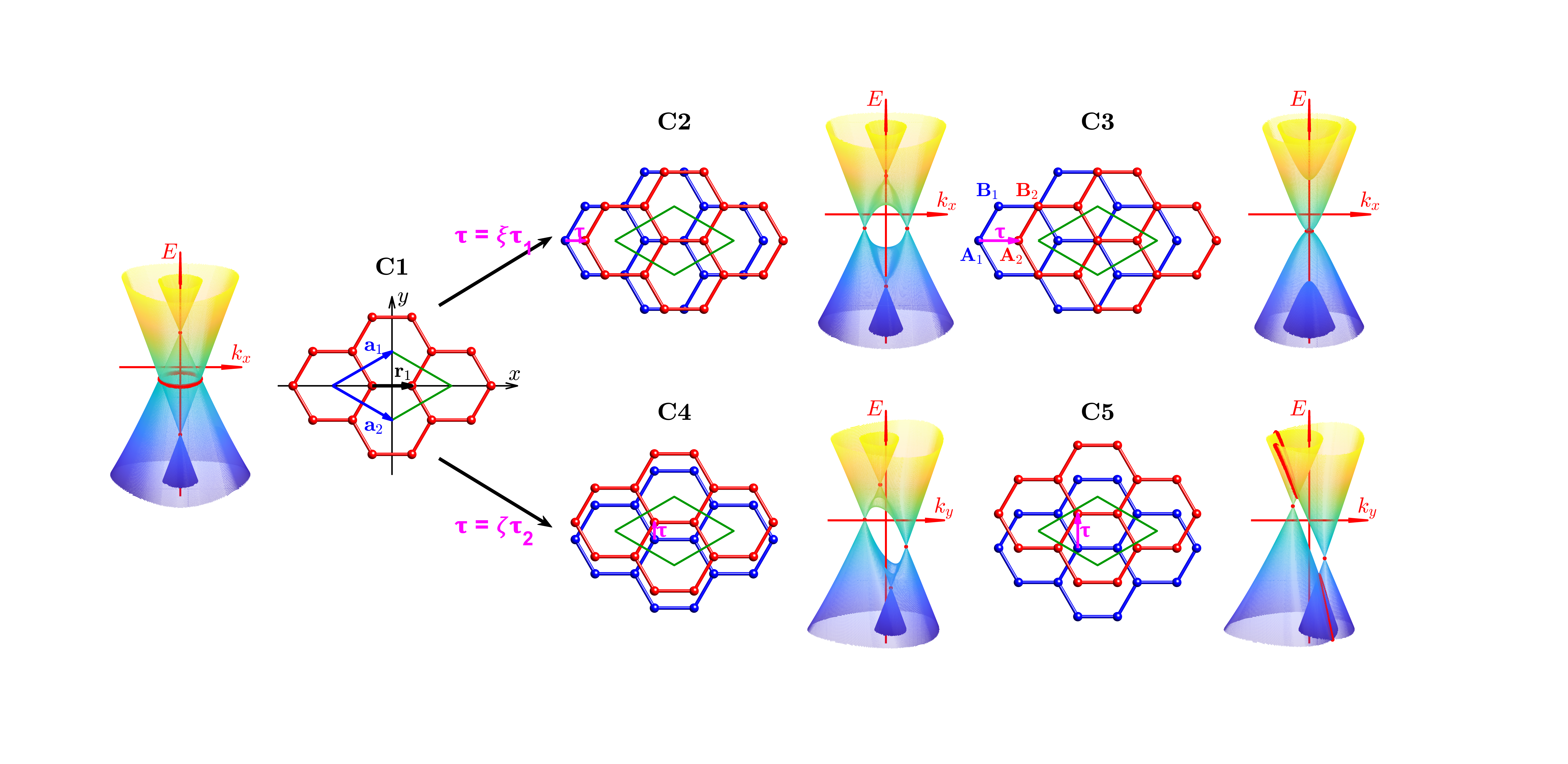}
    \caption{Atomic structures of five representative SBL configurations characterized by typical sliding vectors $\boldsymbol{\tau}$. Configuration C1 corresponds to $\boldsymbol{\tau} = (0,0)$ (AA-stacked), C2 to $\boldsymbol{\tau} = (0.5,0)$, C3 to $\boldsymbol{\tau} = (1,0)$ (AB-stacked), C4 to $\boldsymbol{\tau} = (0,0.5)$, and C5 to $\boldsymbol{\tau} = (0,1)$. The unit cell is indicated by the green rhombus defined by the basis vectors $\mathbf{a}_1$ and $\mathbf{a}_2$. The sliding vector $\boldsymbol{\tau}$ is shown in pink. The corresponding energy surfaces are obtained using Slater-Koster model, which yields $t_2=0.10$ and the following interlayer hopping amplitudes: (C1) $t_u=0.18,\,t_v=t_w=0.08$, (C2) $t_u=t_v=0.14,\,t_w=0.10$, (C3) $t_v=0.18,\,t_u=t_w=0.08$, (C4) $t_u=0.15,\,t_v=t_w=0.12$, and (C5) $t_u=0.10,\,t_v=t_w=0.14$.}
    \label{Fig_1}
\end{figure*}

\subsection{Calculation of Chern numbers}\label{Sec_II.D}
It is worth noting that the introduction of the Haldane terms---characterized by the magnetic phases $\phi_1$ and $\phi_2$---explicitly breaks time-reversal symmetry of the total Hamiltonian. Consequently, the resulting electronic phases of the SBL configurations are naturally classified by the Chern number $C$, a topological invariant computed for the manifold of valence-band states (i.e., the states below the intrinsic Fermi level). 

In the two-dimensional framework considered here, the valence sector comprises two low-energy bands—the first and second valence bands. For every crystal momentum $\mathbf{k}$ in the Brillouin torus $T^{2}$, we define the associated valence manifold as
\begin{equation}
    \mathcal{M}_{V} = \left\{ \sum_{n=1}^2 \alpha_{n\mathbf{k}} |n,\mathbf{k}\rangle \,\middle|\, \alpha_{n\mathbf{k}} \in \mathbb{C}, \sum_{n=1}^2 |\alpha_{n\mathbf{k}}|^2 = 1, \mathbf{k}\in T^2 \right\}.
\end{equation}
This manifold is invariant under $U(2)$ gauge transformations, allowing the use of non-Abelian gauge theory to compute $C$. The Chern number is given by
\begin{equation}\label{Eq21}
    C = \frac{1}{2\pi}\text{Tr} \left\{ \int_{\text{BZ}} F^{\mu\nu}(\mathbf{k})\, dk_\mu dk_\nu \right\},
\end{equation}
where $F^{\mu\nu}(\mathbf{k})$ are the components of the curvature two-form $F(\mathbf{k})$, defined in terms of the connection one-form $A(\mathbf{k})$ as
\begin{equation}\label{Eq22}
    F^{\mu\nu}(\mathbf{k}) = \frac{\partial A^\nu(\mathbf{k})}{\partial k_\mu} - \frac{\partial A^\mu(\mathbf{k})}{\partial k_\nu} - i[A^\mu(\mathbf{k}), A^\nu(\mathbf{k})].
\end{equation}
The connection components $A^\mu(\mathbf{k})$ are $2 \times 2$ matrices with elements
\begin{equation}\label{Eq23}
    A^\mu_{mn}(\mathbf{k}) = i \langle m,\mathbf{k}| \frac{\partial}{\partial k_\mu} |n,\mathbf{k} \rangle, \quad m,n \in \{1,2\}.
\end{equation}

However, computing $C$ from Eq.~(\ref{Eq21}) is numerically inefficient due to (1) the gauge dependence of $A^\mu_{mn}(\mathbf{k})$ and (2) the need for numerical derivatives of second order. An equivalent and more convenient expression is given by
\begin{equation}\label{Eq24}
    C = \frac{1}{2\pi i} \int_{\text{BZ}} \text{Tr} \left\{ P(\mathbf{k}) \left[ \frac{\partial P(\mathbf{k})}{\partial k_\mu}, \frac{\partial P(\mathbf{k})}{\partial k_\nu} \right] \right\} dk_\mu dk_\nu,
\end{equation}
where $P(\mathbf{k}) = \sum_{n=1}^2 |n,\mathbf{k}\rangle \langle n,\mathbf{k}|$ is the projector onto the valence-band subspace.\cite{Bhattacharjee_2021} Since $P(\mathbf{k})$ is gauge invariant and Eq.~(\ref{Eq24}) involves only first-order derivatives, it provides a more stable and efficient method for numerical evaluation. For our two-dimensional systems, the indices $\mu,\nu$ are taken to be $x$ and $y$.

\begin{figure*}
    \centering
    \includegraphics[width=\textwidth]{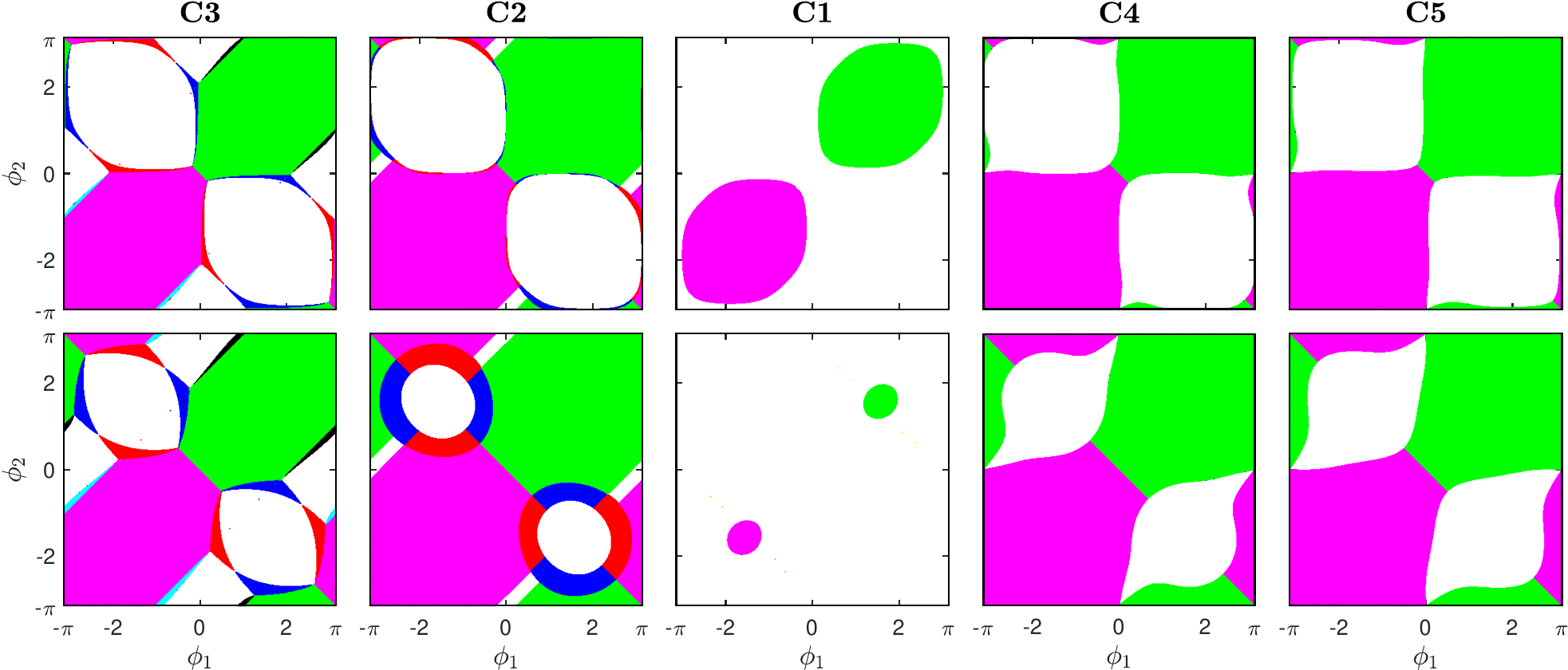}\vspace{0.2cm}
    % The color labels:
    {\fontsize{8}{8}\selectfont\textcolor{black}{\fboxsep=0pt\fbox{\textcolor[HTML]{000000}{$\blacksquare$}} $\,C=-3\;\;$ \fbox{\textcolor[HTML]{00FF00}{$\blacksquare$}} $\,C=-2\;\;$ \fbox{\textcolor[HTML]{0000FF}{$\blacksquare$}} $\,C=-1\;\;$ \fbox{\textcolor[HTML]{FFFFFF}{$\blacksquare$}} $\,C=0\;\;$ \fbox{\textcolor[HTML]{FF0000}{$\blacksquare$}} $\,C=+1\;\;$ \fbox{\textcolor[HTML]{FF00FF}{$\blacksquare$}} $\,C=+2\;\;$ \fbox{\textcolor[HTML]{00FFFF}{$\blacksquare$}} $\,C=+3$}}
    % The caption:
    \caption{Phase diagrams of five SBL configurations computed using Slater-Koster model. The hopping amplitudes in the top row are identical to those in Fig.~\ref{Fig_1}. In the bottom row, all interlayer hopping parameters are uniformly scaled by a factor of $2.8$, resulting in the following values: (C1) $t_u=0.50,\,t_v=t_w=0.22$, (C2) $t_u=t_v=0.39,\,t_w=0.28$, (C3) $t_v=0.50,\,t_u=t_w=0.22$, (C4) $t_u=0.42,\,t_v=t_w=0.34$, and (C5) $t_u=0.28,\,t_v=t_w=0.39$.}
    \label{Fig_2}
\end{figure*}

\section{Results and discussion}\label{Sec_III}
\subsection{Numerical setup and investigation strategy}
Numerical calculations are performed using a rhombic unit cell of the reciprocal space, i.e., the Brillouin torus $T^2$, instead of the hexagonal Brillouin zone, to sample the $\mathbf{k}$ values. This choice simplifies the meshing of the $\mathbf{k}$ grid and ensures the periodicity of the Bloch states, $|n,\mathbf{k}+\mathbf{b}_{1,2}\rangle = |n,\mathbf{k}\rangle$. For each $\mathbf{k}$ point, we carry out exact diagonalization of the Bloch Hamiltonian $H(\mathbf{k})$ to obtain its four eigenvalues and corresponding eigenvectors $\{|n,\mathbf{k}\rangle,\, n=1,\dots,4\}$. Note that the obtained eigenvectors are in a random gauge. To compute the Chern numbers, we employ two complementary approaches: the Fukui--Suzuki method for single-band manifolds,\cite{Fukui_2005} and the multiband method described in Sec.~\ref{Sec_II.D}.\cite{Bhattacharjee_2021} The latter is also applicable to the single-band case.

As a benchmark, we recalculated the phase diagram for a bilayer system with zero interlayer coupling and identical parameters in both layers. The resulting band structure is identical to that of the monolayer Haldane model, exhibiting a twofold degeneracy. The computed $m$--$\phi$ phase diagram reproduces the well-known result that the curves $m=\pm 2\sqrt{3}t_{2}\sin\phi$ define the phase boundaries separating regions with distinct Chern numbers. Specifically, within the range $-2\sqrt{3}t_{2}|\sin\phi|<m<2\sqrt{3}t_{2}|\sin\phi|$, the system exhibits a nontrivial topological insulating phase with $C=1$ for $\phi\in(0,\pi/2)$ and $C=-1$ for $\phi\in(-\pi/2,0)$. Outside this region, the Chern number is $C=0$, indicating a trivial insulator.\cite{Haldane_1988} In the special case of $m=0$, where inversion symmetry is preserved, the system remains topologically nontrivial for all $\phi\neq 0,\pi$, while the bulk gap closes at $\phi=0$ and $\pi$.

We further calculated the Chern numbers of the two valence bands for the bilayer Haldane model in the AB-stacking configuration, where only the vertical interlayer hopping $t_v$ between nearest atoms in adjacent layers is activated. The resulting phase diagram agrees with the results of Mondal \textit{et al.}\ for direct interlayer coupling,\cite{Mondal_2023} and of Bhattacharjee \textit{et al.}\ for staggered coupling,\cite{Bhattacharjee_2021} confirming the accuracy of our implementation and consistency with established results.

Before presenting the main results, we briefly summarize the model parameters and computational setup. The nearest-neighbor hopping amplitude within each atomic layer is taken as the energy unit, and all other parameters are scaled accordingly. Incorporating Haldane-type complex hoppings in both layers introduces, in addition to the interlayer coupling, four intrinsic parameters $(m_{1},m_{2},\phi_{1},\phi_{2})$ that control the opening of the bulk energy gap in all SBL configurations considered. In their absence, i.e., when $m_{1}=m_{2}=\phi_{1}=\phi_{2}=0$, time-reversal symmetry is preserved and all configurations remain gapless.

Figure~\ref{Fig_1} displays the corresponding energy surfaces for five representative SBL configurations in the vicinity of the $\mathbf{K},\mathbf{K}^\prime$ points of the Brillouin torus $T^{2}$. The calculated results reveal pronounced band touchings and crossings, particularly the emergence of small Dirac points near $\mathbf{K},\mathbf{K}^\prime$ for configurations C2, C4, and C5. In configuration C3, the second and third energy bands exhibit overall parabolic dispersions, yet their contact points can evolve into mini Dirac cones, typically known as the wraping of energy surfaces,\cite{Cserti_2007, Predin_2016, Zeng_2017} depending on the magnitude of the skew interlayer couplings.

\begin{figure*}
    \centering
    \includegraphics[width=\linewidth]{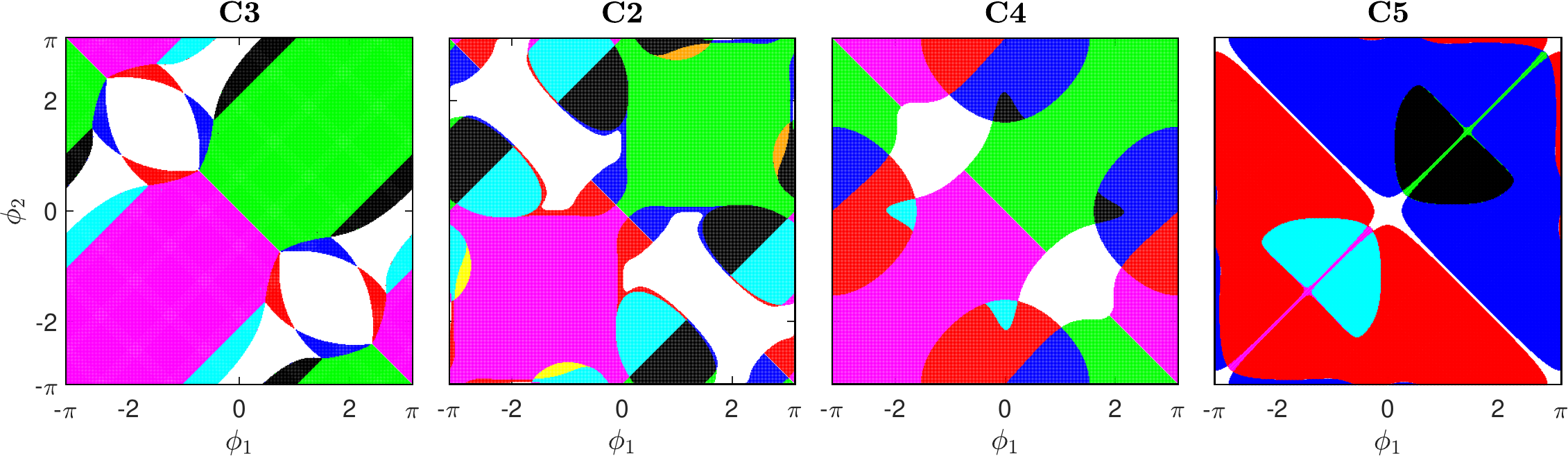}\vspace{0.2cm}
    % The color labels:
{\fontsize{8}{8}\selectfont\textcolor{black}{\fboxsep=0pt \fbox{\textcolor[HTML]{FFA500}{$\blacksquare$}} $\,C=-4\;\;$ \fbox{\textcolor[HTML]{000000}{$\blacksquare$}} $\,C=-3\;\;$ \fbox{\textcolor[HTML]{00FF00}{$\blacksquare$}} $\,C=-2\;\;$ \fbox{\textcolor[HTML]{0000FF}{$\blacksquare$}} $\,C=-1\;\;$ \fbox{\textcolor[HTML]{FFFFFF}{$\blacksquare$}} $\,C=0\;\;$ \fbox{\textcolor[HTML]{FF0000}{$\blacksquare$}} $\,C=+1\;\;$ \fbox{\textcolor[HTML]{FF00FF}{$\blacksquare$}} $\,C=+2\;\;$ \fbox{\textcolor[HTML]{00FFFF}{$\blacksquare$}} $\,C=+3\;\;$ \fbox{\textcolor[HTML]{FFFF00}{$\blacksquare$}} $\,C=+4$}}
    \caption{Chern number diagrams for four SBL configurations C2, C3, C4, and C5 as functions of the phases $\phi_1$ and $\phi_2$. The C1 (AA-stacked) configuration does not exhibit a phase with $|C|>2$ and is therefore not included. The values of hopping amplitudes are taken to be (C2) $t_2=0.56,\,t_u=0.90,\,t_v=0.50,\,t_w=0.50$, (C3) $t_2=0.10,\,t_u=0.08,\,t_v=0.70,\,t_w=0.60$, (C4) $t_2=0.30,\,t_u=0.90,\,t_v=t_w=0.70$, and (C5) $t_2=0.30,\,t_u=0.50,\,t_v=t_w=0.20$.}
    \label{Fig_3}
\end{figure*}

To reduce the dimensionality of the parameter space in the subsequent analysis, we fix $m_{1}=m_{2}=0$ and treat $(\phi_{1},\phi_{2})$ as the only tunable parameters. This restriction isolates the effects of time-reversal-symmetry breaking while preserving inversion symmetry, yielding a clear and tractable phase diagram in the $(\phi_{1},\phi_{2})$ plane. Moreover, eliminating the mass terms emphasizes the role of interlayer coupling—particularly the skew hoppings—in shaping the system’s topological character without the confounding influence of layer-specific mass gaps. This parameter choice follows the standard conventions adopted in previous multilayer Haldane studies,\cite{Mondal_2023,Mannai_2023,Bhattacharjee_2021,Chen_2011,Bhattacharjee_2021} thereby allowing direct comparison with earlier results.

Rather than scanning the entire parameter space, we identify topological phase boundaries by monitoring the Chern numbers of the manifolds spanned by the first and second energy bands. An integer-valued Chern number indicates that the corresponding manifold is well defined and isolated from the others. In contrast, irregular or fluctuating values signal that the manifold is ill defined, typically due to degeneracies arising from band crossings with the third or fourth bands. The same analysis is applied at the individual-band level. Using this approach, we find that in all SBL configurations, the two lowest-energy bands remain separated by interlayer coupling, whereas the gap between the second and third bands depends sensitively on $(\phi_1,\phi_2)$. We emphasize, however, that such band separation does not necessarily imply a global insulating gap between valence and conduction bands—it merely ensures the local isolation of the valence manifold $\mathcal{M}_V$, which is sufficient for defining and computing its Chern number. The physical implications of these findings are discussed in the following sections.

\subsection{$\phi_1$--$\phi_2$ phase diagram}
Figure~\ref{Fig_2} presents the Chern numbers of the full valence manifold $\mathcal{M}_V$ as functions of the phase parameters $\phi_1$ and $\phi_2$ for the five SBL configurations. The interlayer coupling amplitudes are varied to assess its impact on the emergence and stability of topological phases. In the absence of interlayer hopping, the bilayer system reduces to a trivial superposition of two single-layer Haldane models, giving $C=+2$ in the first quadrant, $C=-2$ in the third quadrant, and $C=0$ in the remaining two quadrants (not shown). Introducing interlayer coupling substantially modifies this structure. For the AA configuration, the $C=\pm2$ regions evolve into two oval-shaped domains that contract with increasing interlayer hopping strengths and eventually disappear once the coupling becomes sufficiently strong.

For the other stacking geometries, the phase diagrams display richer and more intricate structures, with interwoven regions of $C=0$, $\pm1$, and $\pm2$. Configurations C4 and C5 exhibit qualitatively similar patterns, consistent with their identical (i.e., absent) symmetry constraints. Configurations C2 and C3, while related, differ quantitatively due to their distinct symmetry properties; the AB-stacked configuration (C3) possesses a higher degree of symmetry than C2. Increasing the interlayer coupling systematically shrinks the $C=0$ regions, reflecting enhanced interlayer-induced band mixing. Remarkably, for configuration C3, narrow pockets with $C=\pm3$ appear, as highlighted by the black and cyan domains in Fig.~\ref{Fig_2}. These regions are extremely small and require fine-tuning of parameters, indicating that the corresponding high-Chern-number phases are delicate.

Figure~\ref{Fig_3} further highlights the appearance of topological phases with $C=\pm3$ in configurations C2, C3, C4, and C5; configuration C1 is omitted because no such phases were observed. (The color scale representing different Chern numbers is defined in the figure caption.) In addition to the $|C|=3$ phases, configuration C2 also hosts small pockets with $|C|=4$, which is notable because it exceeds the sum of the Chern numbers of the two independent monolayers.

\begin{figure}[t!]
     \centering
     \includegraphics[width=\linewidth]{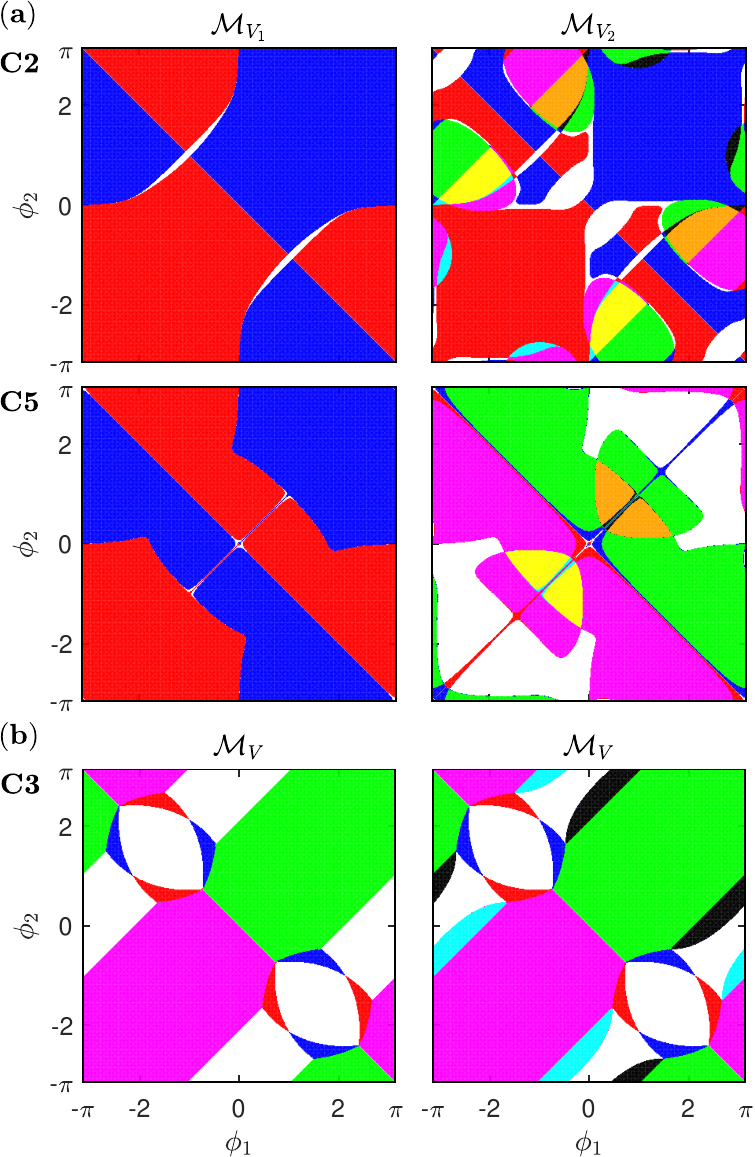}
     \caption{(a) Phase diagrams of two valence manifolds $\mathcal{M}_{V_1}$ and $\mathcal{M}_{V_2}$ corresponding to the two configurations C2 and C5 presented in FIG.~\ref{Fig_3}. (b) Phase diagrams of the valence manifold $\mathcal{M}_V$ for a representative configuration C3 computed in the two cases: (left panel) all skew hoppings absent with $t_v=0.70,\;t_u=t_w=0$, and (right panel) one skew hopping present with $t_u=0,\;t_v=0.70,\;t_w=0.60$. The color codes here are identical to those in FIG.~\ref{Fig_3}.}
     \label{Fig_4}
\end{figure}

To elucidate the mechanisms responsible for the emergence of high–Chern-number phases, we separately evaluate the Chern numbers associated with the two valence-band submanifolds, $\mathcal{M}_{V_1}$ and $\mathcal{M}_{V_2}$. Representative phase diagrams for configurations C2 and C5 are displayed in Fig.~\ref{Fig_4}(a). The topology of $\mathcal{M}_{V_1}$ is relatively simple, exhibiting only $C=0$ (for C2) and $C=\pm 1$ regions. In contrast, the phase diagram of $\mathcal{M}_{V_2}$ is considerably richer: although $C=\pm 1$ regions dominate, isolated pockets with $C=\pm 3$ and even $C=\pm 4$ appear near $(0,\pm\pi)$ and $(\pm\pi,0)$. This enhanced complexity originates from strong interlayer hybridization involving the second and third bands. While the structure of $\mathcal{M}_{V_1}$ differs from that of a single-layer Haldane model, it nonetheless retains characteristic signatures of the four-band structure induced by interlayer coupling.

A central observation is that, in configurations C2, C4, and C5, the skew interlayer hoppings dominate the interlayer hybridization, whereas the vertical hopping channel plays a significant role only in configurations C1 and C3. To highlight the impact of the skew hoppings on stabilizing phases with $|C|>2$, Fig.~\ref{Fig_4}(b) contrasts the Chern phase diagrams obtained with only the vertical interlayer coupling retained (left panel, fully consistent with Ref.~\onlinecite{Mondal_2023}, where the AB-stacked configuration was studied in the simplified limit retaining only the vertical interlayer coupling $v_{\mathbf{k}}$) and with the momentum-dependent skew coupling $w_{\mathbf{k}}$ reinstated (right panel). In the absence of skew hoppings, the maximal Chern number is restricted to $|C|=2$. Upon restoring $w_{\mathbf{k}}$, however, sizable regions exhibiting $|C|=3$ phases develop. This behavior aligns with the emergence of high–Chern-number phases ($|C|\ge 3$) in the SBL configurations C2, C4, and C5. Notably, the extent of these large-$|C|$ regions depends sensitively on the strength of the skew hoppings: if the skew terms are insufficiently strong, the $|C|>2$ regions shrink and may disappear entirely. This demonstrates that robust skew interlayer hybridization is essential for enabling and stabilizing the large-Chern-number phases.

\begin{figure}
     \centering
     \includegraphics[clip,trim=8.75cm 0cm 12.85cm 0cm,width=\linewidth]{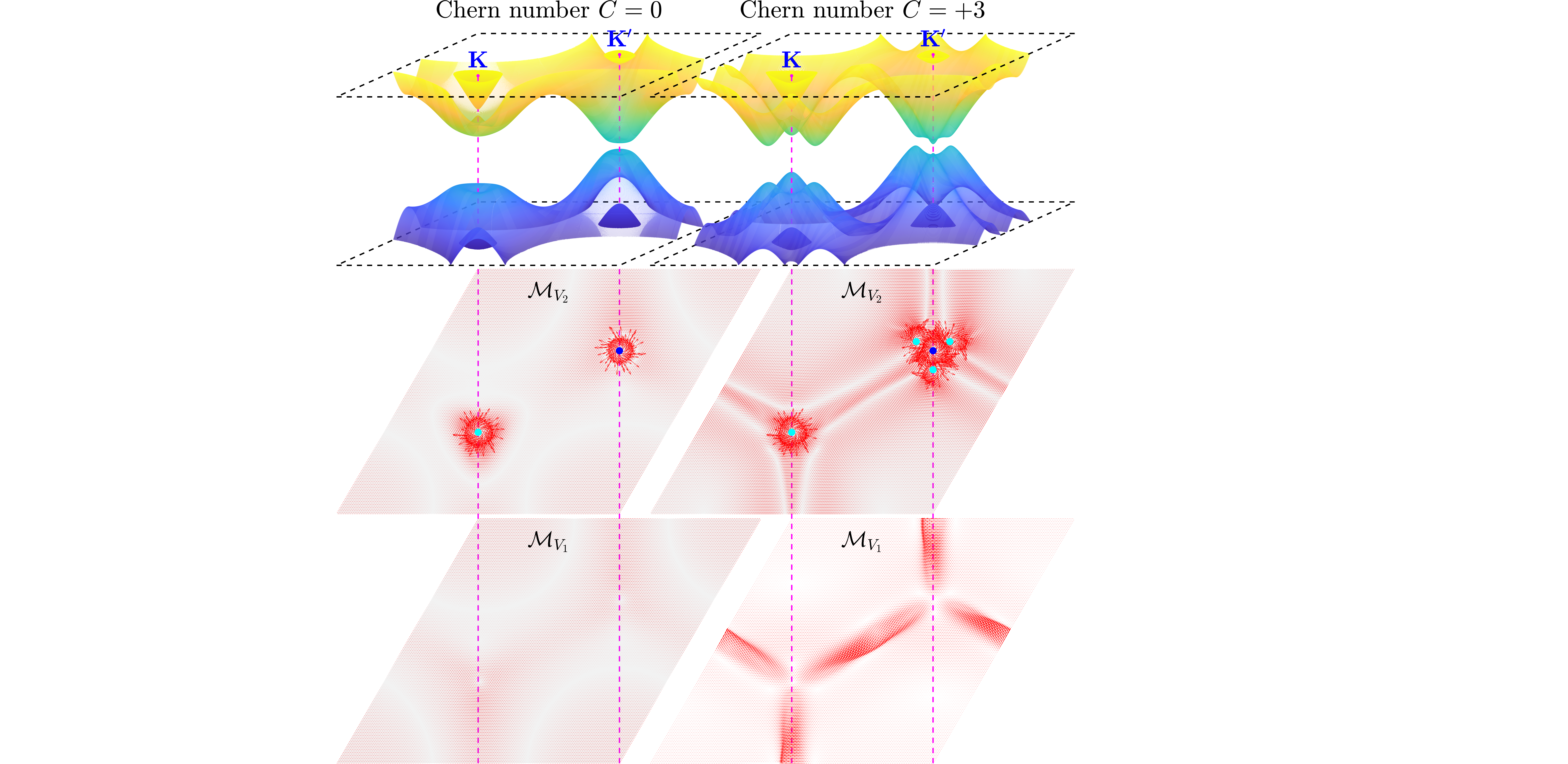}
     \caption{Energy surfaces and Berry connection fields, $\mathbf{A}(\mathbf{k})$, of valence manifolds $\mathcal{M}_{V_1}$ and $\mathcal{M}_{V_2}$ plotted for a representative configuration C3 in the two cases: (left column) all skew hoppings absent with $t_v=0.70,\;t_u=t_w=0$, and (right column) one skew hopping present with $t_u=0,\;t_v=0.70,\;t_w=0.60$. The Haldane phases in both cases are $(\phi_1,\phi_2)=(-2.60,-0.20)$. The cyan and blue dots indicate the positive and negative signs of Berry curvature, respectively, computed by integrating over vortex regions of $\mathbf{A}(\mathbf{k})$. The black and magenta dashed lines are plotted to denote the Brillouin zone and to align $\mathbf{K},\mathbf{K}^\prime$ valleys, respectively.}
     \label{Fig_5}
 \end{figure}

Figure~\ref{Fig_5} illustrates the energy surfaces of the C3 configuration, computed at a representative point $(\phi_{1},\phi_{2})$ that supports a $C=+3$ phase when all interlayer couplings are active. The upper-left panel shows the band structure obtained when the skew hoppings are switched off. In this limit, the four bands remain well separated and exhibit simple dispersions; the gap between the second and third bands is relatively large at the $\mathbf{K}$ point and considerably narrower at $\mathbf{K}^\prime$. Consistent with this trivial topology, the resulting Chern number is $C=0$. When the skew hoppings are reinstated, however, the band surfaces acquire substantially more intricate features, especially for the second and third bands. A narrow but finite global gap separating these two bands persists, yet the first (fourth) and second (third) bands are no longer fully isolated: local band merging occurs near the $\mathbf{K}^\prime$ point. Notably, the second band develops a volcano-like structure characterized by a deep central minimum surrounded by three elevated ridges, while the third band exhibits a complementary structure in which the roles of $\mathbf{K}$ and $\mathbf{K}^\prime$ are interchanged (see the upper-right panel).

To further characterize these phases, we compute the Berry connection across the Brillouin zone using parallel-transport gauge fixing to ensure a smooth choice of phases. For the $C=0$ case, Fig.~\ref{Fig_5} shows that the Berry connection $\mathbf{A}(\mathbf{k})$ associated with $\mathcal{M}_{V_1}$ is nearly negligible throughout the Brillouin zone, whereas $\mathcal{M}_{V_2}$ hosts two vortices of opposite chirality located near the $\mathbf{K}$ and $\mathbf{K}^\prime$ points. The corresponding Berry curvature is sharply localized at these vortices and integrates to zero over the Brillouin-zone torus $T^{2}$, consistent with the trivial topology of the manifold. In contrast, for the $C=+3$ phase, the Berry-connection field of $\mathcal{M}_{V_1}$ does not form isolated vortex structures; instead, it develops extended anisotropic flow patterns aligned with the high-symmetry directions of $T^{2}$. The manifold $\mathcal{M}_{V_2}$, however, exhibits three pronounced vortices clustered near the $\mathbf{K}^\prime$ point. Taken together, the Berry-connection field over the full torus contains five vortices: four with the same chirality and one with the opposite. The Berry curvature is strongly concentrated near these vortex centers, and its integral yields a total Chern number of $+3$, as required. This analysis demonstrates explicitly how the skew interlayer hoppings hybridize states between the two layers and enhance the twisting of the valence and conduction manifolds, thereby enabling the emergence of topological phases with Chern numbers reaching or exceeding~$3$.
 
\begin{figure}
     \centering
     \includegraphics[clip,trim=12.95cm 0.1cm 13.12cm 0.15cm,width=\linewidth]{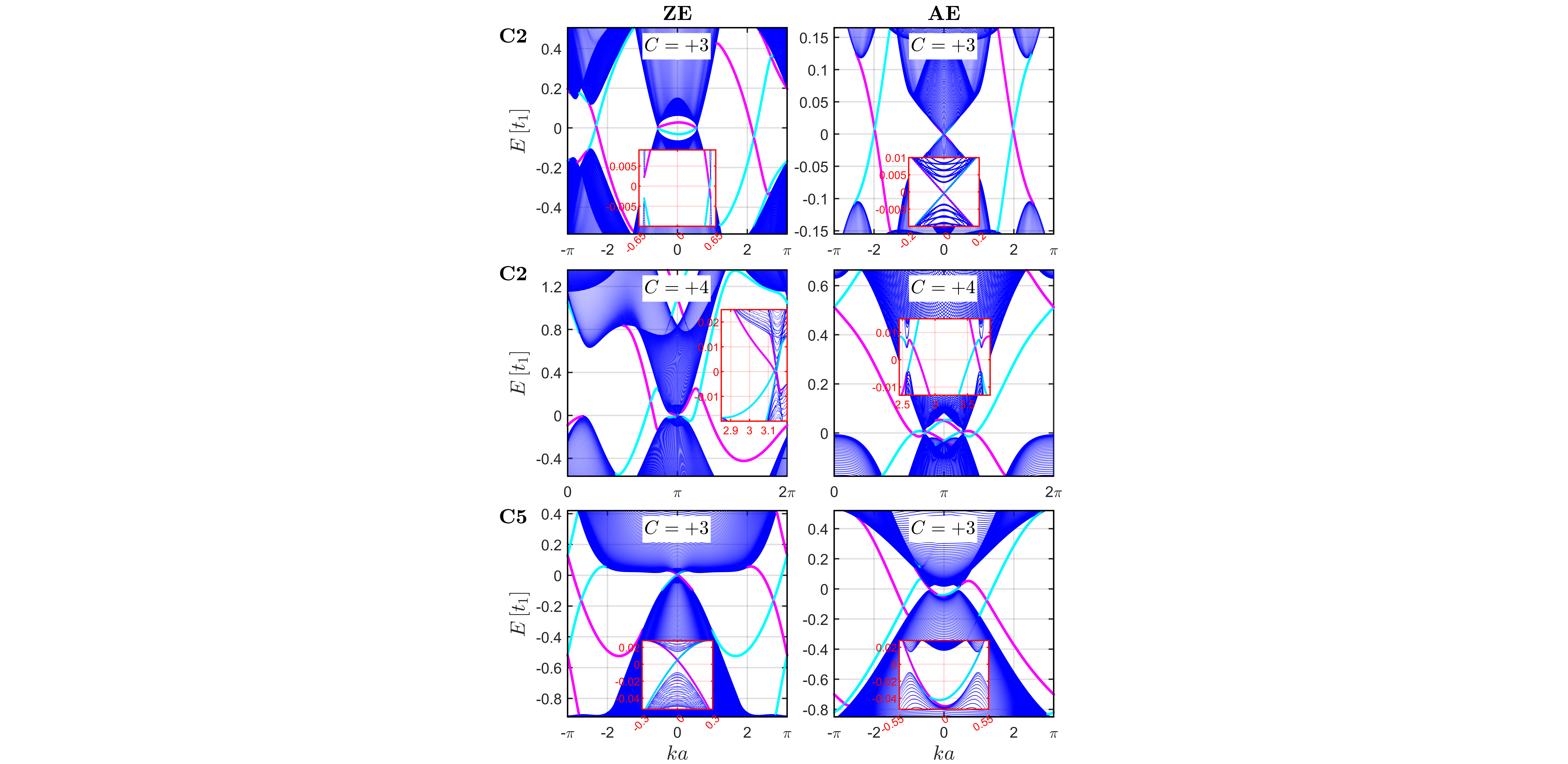}\\
     \includegraphics[,width=\linewidth]{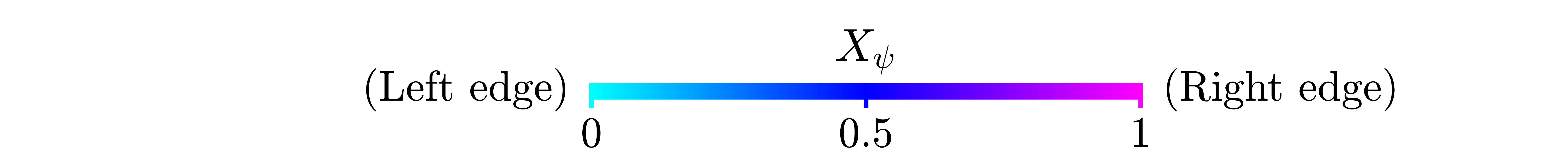}
     \caption{Band structures of two nanoribbon configurations with zigzag and armchair edges plotted for two representative SBL configurations C2 and C5. The values of parameters are sellected as: (C2) with $C=+3$: $t_2=0.56,\,t_u=0.20,\,t_v=0.73,\,t_w=0.38,\,(\phi_1,\phi_2)=(-2.72,-0.43)$; (C2) with $C=+4$: $t_2=0.56,\,t_u=0.80,\,t_v=0.01,\,t_w=0.90,\,(\phi_1,\phi_2)=(-2.34,-0.85)$; and (C5) with $C=+3$: $t_2=0.30,\,t_u=0.50,\,t_v=t_w=0.20,\,(\phi_1,\phi_2)=(-1.80,-0.64)$. The ribbon width is specified by $N_\text{arm}=1627$ for AE ribbon of configuration C2 with $C=+3$ (top-right panel) and $N_\text{zig}=N_\text{arm}=600$ for the other configurations. The color scale represents the quantity $X_\psi$, quantifying the spatial localization of eigenstates.}
     \label{Fig_6}
\end{figure}

\subsection{Chiral edge modes}
The Chern phase diagrams discussed above do not, by themselves, identify the regions of parameter space in which the bilayer system realizes a genuine insulating phase. Although the second and third bands may remain nondegenerate throughout the Brillouin zone, the maximum of the former may still lie above the minimum of the latter, thereby preventing the opening of a bulk gap. Such situations—frequently encountered in configurations C4 and C5—allow one to assign a Chern number to the corresponding isolated band manifold, but the resulting value does not correspond to a physically realized insulating state. To establish the presence of topologically nontrivial insulating phases, it is therefore necessary to examine the electronic spectra directly and to verify (i) the existence of a finite bulk gap separating the valence and conduction manifolds, and (ii) the presence of chiral edge-localized states traversing this gap. To this end, we compute the band structures of nanoribbons with sufficiently large widths to suppress quantum-confinement effects and approximate the thermodynamic limit.

The resulting ribbon spectra display substantial variation across stacking configurations and parameter regimes. For armchair and zigzag ribbons constructed from bilayer models with sliding vector $\boldsymbol{\tau}=\xi\boldsymbol{\tau}_1$, we generally find a narrow but finite separation between the valence (bands 1 and 2) and conduction (bands 3 and 4) manifolds, accompanied by chiral edge states that cross the gap. By contrast, ribbons based on configurations with $\boldsymbol{\tau}=\zeta\boldsymbol{\tau}_2$ typically lack such a bulk gap, making it more challenging to stabilize insulating phases. In certain parameter windows, however, modest tuning of the interlayer hopping amplitudes yields a small but finite bulk gap.

To characterize the spatial localization of the edge states, we employ a color-mapping scheme based on the quantity $X_\psi \!\in [0,1]$,
\begin{equation}
    X_\psi = (1-\sigma_r)\langle j\rangle_r+0.5\sigma_r,
\end{equation}
where
\begin{eqnarray}
    && \sigma_r = \frac{\sigma}{\sigma_\text{max}}, \nonumber\\
    && \sigma =\sqrt{\langle(j-\langle j\rangle)^2\rangle}, \nonumber\\
    && \sigma_\text{max} = \sqrt{\frac{1}{N}\sum_{j=1}^{N}(j-\langle j\rangle)^2}, \nonumber\\
    && \langle j\rangle_r = \frac{1}{N}\langle j\rangle, \nonumber\\
    && \langle j\rangle = \sum_{j=1}^{N} j\,P_j(n,k), \nonumber\\
    && P_j(n,k) = |\langle R_j|\psi_{n,k}\rangle|^2.
\end{eqnarray}
with $N=N_{A/Z}$. In this convention, $X_\psi \!\to\! 0$ ($X_\psi \!\to\! 1$) indicates a state localized near the left (right) ribbon edge, whereas $\sigma_r$ measures its spatial spread. When $\sigma_r \!\to\! 1$ and $X_\psi \!\to\! 0.5$, the state is fully delocalized and resides in the bulk.

Using this analysis, we confirm the topological insulating phases predicted by the Chern phase diagrams—specifically, those regions in which a bulk gap coexists with chiral edge states whose number equals the Chern number of the occupied manifold. The cases $C=\pm1$ and $C=\pm2$ are readily verified owing to their relatively large bulk gaps. By contrast, the $C=\pm3$ phases are considerably more delicate: they generally exhibit much smaller gaps and occur only within narrow regions of the $(\phi_1,\phi_2)$ parameter space, requiring careful tuning of the interlayer hopping amplitudes.

Figure~\ref{Fig_6} presents representative ribbon spectra for configurations C2 and C5. For both zigzag and armchair terminations with widths $W\simeq 74$--$200$ nm, parameter sets corresponding to the $C=\pm3$ phase in configuration~C2 yield a well-defined bulk gap of approximately $25\,\mathrm{meV}$, within which three pairs of chiral edge modes connect the valence and conduction manifolds. In addition to the $C=\pm3$ phase, the phase diagram in Fig.~\ref{Fig_3} also contains small regions predicting insulating behavior with $C=\pm4$. As noted earlier, the phase diagram only indicates where the valence manifold $\mathcal{M}_V$ is well defined; it does not guarantee the presence of a global bulk gap. We therefore use parameter sets extracted from these $C=\pm4$ regions as a guide and subsequently adjust the interlayer hopping amplitudes to open a bulk gap. The resulting ribbon spectra reveal that four pairs of chiral edge modes are clearly visible despite the smallness of the bulk gaps. These features—robust for both zigzag and armchair terminations—are fully consistent with the bulk--boundary correspondence and confirm the existence of insulating phases with Chern number $C=\pm3$ and $C=\pm4$.

For configuration C5, the energy surfaces frequently exhibit apparent band-touching points between the second and third bands. To determine whether these correspond to genuine degeneracies, we evaluate the Berry phase accumulated along a closed loop encircling each candidate point. The resulting phases deviate from integer multiples of $\pi$, demonstrating that these crossings are accidental and do not correspond to Dirac points. Upon moderate adjustment of the interlayer hopping amplitudes, a small bulk gap of a few tens of meV opens, within which three pairs of chiral edge states clearly appear, as shown in the lower panels of Fig.~\ref{Fig_6}. This again confirms the existence of an insulating phase with Chern number $C=\pm3$.

\subsection{Discussion}\label{Sec_III.D}
Realizing QAH phases with large Chern numbers in real materials remains a formidable challenge, even though such phases are, in principle, classified by an integer ($\mathbb{Z}$) topological invariant. The difficulty lies not in the classification itself,\cite{Wan_2025} but in designing realistic systems that host multiple chiral edge channels protected by sizable bulk gaps. Here we discuss the experimental relevance and physical implications of the approach proposed in this work.

This study examines bilayer systems composed of two atomically thin hexagonal lattices, such as graphene-based structures, where the interlayer coupling is treated comprehensively. While the Bernal-stacked configuration has been widely studied,\cite{Bhattacharjee_2021,Mondal_2023} we emphasize the broader impact of symmetry-controlled hybridization by explicitly including skew hoppings beyond the purely vertical terms. The tight-binding parametrization follows Slater--Koster rules, ensuring a direct correspondence between hopping amplitudes, orbital overlaps, and atomic geometry. This microscopic foundation anchors the model to realistic atomic lattices, unlike phenomenological constructions involving long-range hoppings of unclear physical origin. The dominant interlayer processes remain short-ranged, with smoothly varying magnitudes and phases that can be experimentally tuned through relative translation, twist angle, interlayer spacing, or pressure.

The complex next-nearest-neighbor hoppings within each layer serve as the mechanism for breaking time-reversal symmetry and opening topological gaps. Although an idealized Haldane Hamiltonian with prescribed complex phases may not exist literally in a specific material, equivalent physics has been realized in magnetically doped topological insulators\cite{Kim_2017} and cold-atom systems.\cite{Jotzu_2014} Moreover, recent optical-lattice experiments have demonstrated that complex hoppings can be induced in hBN using counter-rotating laser beams.\cite{Mitra_2024} These advances confirm that the Haldane framework provides a compact, physically grounded model, while the present Slater--Koster formulation preserves its atomic realism.

Compared with other approaches for achieving high–Chern-number phases, the bilayer sliding architecture offers distinct advantages. The Haldane-type complex hopping mechanism does not involve spin or magnetic degrees of freedom, making it suitable for two-dimensional materials composed of light, nonmagnetic elements. In contrast, models requiring fine-tuned long-range hoppings or extremely flat bands are typically fragile against disorder. By exploiting symmetry-governed hybridization between local orbitals, our framework provides a more robust and controllable route to realizing nonadditive, symmetry-controlled topological phases driven by interlayer coupling.

Taken together, our findings may pave a practical route toward realizing high–Chern-number quantum anomalous Hall phases in van der Waals heterostructures. By tuning geometric parameters such as interlayer spacing, lateral displacement, and the phases of complex hoppings, interlayer Chern-phase interactions can be precisely controlled. This framework bridges idealized model Hamiltonians and experimentally accessible materials, offering a versatile platform for designing and tuning complex QAH phenomena.

\section{Conclusion}\label{Sec_IV}
In this work, we have systematically explored how interlayer coupling in bilayer hexagonal Haldane lattices gives rise to quantum anomalous Hall phases with large Chern numbers. By incorporating both vertical and skew interlayer hoppings, we demonstrated that relative sliding between the layers shifts the Dirac cones in momentum space, generating gapless points at generic low-symmetry $\mathbf{k}$ locations. The inclusion of complex next-nearest-neighbor hoppings characteristic of the Haldane model opens gaps at these points, giving rise to a series of QAH phases with Chern numbers exceeding two. Our analysis shows that in stacked systems, the total Chern number is not a simple sum of the single-layer contributions but emerges from the nontrivial hybridization of topological manifolds in the coupled layers.

The resulting $(\phi_1,\phi_2)$ phase diagrams reveal regions with $|C|=3$ and $|C|=4$, confirmed by edge-state calculations showing multiple chiral modes consistent with the bulk topology. We have highlighted the essential role of skew interlayer hoppings in stabilizing these high--Chern-number phases. Furthermore, the bilayer sliding architecture provides a realistic and controllable route to engineering QAH states, with tunable interlayer spacing and lateral displacement. Building on the discussion in Sec.~\ref{Sec_III.D}, these topological phases could, in principle, be realized experimentally in van der Waals heterostructures or hexagonal lattices such as hBN via high-frequency laser driving or other mechanisms that effectively implement complex hoppings.

Overall, our results establish a versatile framework for achieving and manipulating non-additive, symmetry-controlled topological phases in two-dimensional bilayer systems, bridging the gap between theoretical models and experimentally accessible quantum anomalous Hall platforms.\\

\section*{ACKNOWLEDGMENT}
H. Minh Lam was funded by the Master, PhD Scholarship Programme of Vingroup Innovation Foundation (VINIF), code VINIF.2024.TS.014.

%\newpage
\bibliography{bibliography}

\end{document}